\documentclass[aps,prd,twocolumn,groupedaddress,showpacs,nofootinbib]{revtex4-1}
\usepackage{amsmath,amssymb,graphicx}

\usepackage{graphicx}
\usepackage{textcomp}
\usepackage{color}

\def\theta{\vartheta}

\newcommand{\be}{\begin{equation}}
\newcommand{\ee}{\end{equation}}
\newcommand{\ba}{\begin{eqnarray}}
\newcommand{\ea}{\end{eqnarray}}
\newcommand{\lsim}   {\mathrel{\mathop{\kern 0pt \rlap
  {\raise.2ex\hbox{$<$}}}
  \lower.9ex\hbox{\kern-.190em $\sim$}}}
\newcommand{\gsim}   {\mathrel{\mathop{\kern 0pt \rlap
  {\raise.2ex\hbox{$>$}}}
  \lower.9ex\hbox{\kern-.190em $\sim$}}}

\begin{document}

\title{Propagation of Superluminal PeV IceCube Neutrinos: A High Energy Spectral Cutoff or New Constraints on Lorentz Invariance Violation}

\author{Floyd W. Stecker}
\affiliation{Astrophysics Science Division\\NASA Goddard Space Flight Center\\Greenbelt, MD 20771, USA } 

\author{Sean T. Scully}
\affiliation{Department  of Physics\\ James Madison University\\ Harrisonburg, VA 22807}

\begin{abstract}
The IceCube observation of cosmic neutrinos with $E_{\nu} > 60$ TeV, most of which are likely of extragalactic origin, allows one to severely constrain Lorentz invariance violation (LIV) in the neutrino sector, allowing for the possible existence of superluminal neutrinos. The subsequent neutrino energy loss by vacuum $e^+e^-$ pair emission (VPE) is strongly dependent on the strength of LIV. In this paper we explore the physics and cosmology of superluminal neutrino propagation. We consider a conservative scenario for the redshift distribution of neutrino sources. Then by propagating a generic neutrino spectrum, using Monte Carlo techniques to take account of energy losses from both VPE and redshifting, we obtain the best present constraints on LIV parameters involving neutrinos. We find that $\delta_{\nu e } = \delta_{\nu} - \delta_e \le 5.2 \times 10^{-21}$. Taking $\delta_e \le 5 \times 10^{-21}$, we then obtain an upper limit on the superluminal velocity fraction for neutrinos alone of $1.0 \times 10^{-20}$. Interestingly, by taking $\delta_{\nu e} = 5.2 \times 10^{-21}$, we obtain a cutoff in the predicted neutrino spectrum above 2 PeV that is consistent with the lack of observed neutrinos at those energies, and particularly at the Glashow resonance energy of 6.3 PeV. Thus, such a cutoff could be the result of neutrinos being slightly superluminal, with $\delta_{\nu}$ being $(0.5 \ {\rm to} \ 1.0) \times 10^{-20}$.
\end{abstract}


\maketitle

\section{Introduction}

The possible existence of superluminal neutrinos as a consequence of Lorentz invariance violation
(LIV) was brought to the attention of the physics community by their apparent observation 
~\cite{ad11}. Shortly thereafter, Cohen and Glashow~\cite{cg11} presented a powerful theoretical argument against the results in Ref.~\cite{ad11}. Their argument was based on the implication that these neutrinos would rapidly lose energy by the dominant energy loss channel of vacuum electron-positron pair emission (VPE), i.e., $\nu \to \nu \,e^+\, e^-$. Eventually, the results in Ref.~\cite{ad11} were retracted~\cite{ad12}. (See also Ref.~\cite{ad13}).

The Ice Cube collaboration has recently reported the observation of 37 extraterrestrial neutrinos with energy above $\sim$ 60 TeV, giving a cosmic neutrino signal $5.7\sigma$ above the atmospheric background~\cite{wh14}. This is significant evidence for a neutrino flux of cosmic origin, above that produced by atmospheric cosmic-ray secondaries~\cite{aa13a}.%,aa13b,la13}. 
The very existence of PeV neutrinos has been used to place strong constraints on LIV in the neutrino sector ~\cite{st14},~\cite{bo13}.%,di14}.

\section{Cosmic High Energy Neutrinos}

There are four indications that the cosmic neutrinos observed by IceCube are extragalactic in origin~\cite{st14}: (1) The celestial distribution of the 37 reported cosmic events is consistent with isotropy, with no significant enhancement in the galactic plane~\cite{wh14}, although it has been argued that a subset of these events might be of galactic origin~\cite{ra13a}.%,ah13}. 
(2) A possible cutoff  in the energy spectrum of these neutrinos may be indicative of photopion production followed by pion decay~\cite{wi13} such as expected in AGN cores~\cite{st91},
GRBs~\cite{li13}, or intergalactic interactions~\cite{ka13}. (AGN jets have also been looked at, but there may be difficulties with the jet models~\cite{mu13}. 
Neutrinos from starburst galaxies are discussed in Section VI.)  (3) The diffuse galactic neutrino flux~\cite{st79} is expected to be well below that observed by  Ice Cube. (4) At least one of the $\sim$1 PeV neutrinos observed by IceCube (dubbed "Ernie") came from a direction off of the galactic plane. 

An upper limit for the difference between putative superluminal neutrino and electron velocities of $\delta_{\nu e} \equiv \delta_{\nu} - \delta_e \le \sim 5.6 \times 10^{-19}$ was previously derived by one of us, confirming that the observed PeV neutrinos could have reached Earth from extragalactic sources. After obtaining an upper limit on the superluminal electron velocity of $\delta_{e}\equiv v_{e} - 1 \le \ \sim 5 \times 10^{-21}$, an upper limit of $\delta_{\nu} \equiv v_{\nu} - 1 \le \ \sim  5.6 \times 10^{-19}$ was derived from one of the $\sim$PeV neutrino events~\cite{st14}. (Here $c = 1$ and $\delta_{\nu} = - \mathaccent'27 c^{(4)}$ in the standard model extension (SME) effective field theory framework for describing the effects of LIV and CPT violation~\cite{ck98}). This previous limit allows for the possibility that minimally superluminal neutrinos can propagate over large distances from extragalactic sources such as active galactic nuclei (AGN) and $\gamma$-ray bursts (GRB), while undergoing energy losses by VPE.

\begin{figure}[t]
{\includegraphics[scale=0.58]{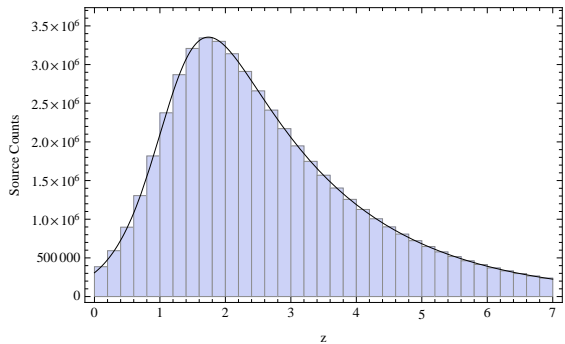}}
\caption{Binned number of neutrinos used in our Monte Carlo runs sampled from the star formation rate distribution of Ref.~\cite{be13}.}
\label{sfr}
\end{figure}

Given that neutrinos detected by IceCube are extragalactic, cosmological effects should be taken into account in deriving new LIV constraints. The reasons are straightforward. As opposed to the extinction of high energy extragalactic photons through electromagnetic interactions~\cite{st92}, neutrinos survive from all redshifts because they only interact weakly. We thus consider here a scenario where the neutrino sources have a redshift distribution that follows that of the star formation rate~\cite{be13} (see Figure~\ref{sfr}), as appears to be roughly the case for both active galactic nuclei and $\gamma$-ray bursts. Since the universe is transparent to neutrinos, most of the cosmic PeV neutrinos will come from sources at redshifts between $\sim$0.5 and $\sim$2~\cite{st91}. Therefore neither energy losses by redshifting of neutrinos nor the cosmological $\Lambda$CDM redshift-distance relation can be neglected in our calculations. 

\section{Neutrino Energy Losses}
 
We again note the definitions $\delta_{\nu e} = \delta_{\nu} - \delta_e, \delta_{\nu} = v_{\nu} - 1$ and $\delta_{e} = v_{e} - 1$. The $v$'s here are to be understood to be the maximum attainable velocities of the neutrinos and electrons respectively. (N.B.: The definition of $\delta$ used here is {\it half} that used in Refs.~\cite{cg11} and~\cite{co99}, but is consistent with that used in Ref.~\cite{sg01}.) For $\delta_{\nu} \ge \delta_e \ge 0$ and defining $\delta_{\nu e} \equiv \delta_{\nu} - \delta_e$, the VPE process $\nu \to \nu\, e^+\, e^-$ is kinematically allowed  provided that~\cite{co99,sg01} 
\begin{equation}
E_{\nu} \ge m_e\sqrt{2/\delta_{\nu e}}
\label{threshold}
\end{equation} 
 
The decay width for the VPE process, $\nu \to \nu \,e^+\, e^-$, is given by~\cite{cg11}
\begin{equation}
\Gamma = \frac{1}{14}\frac{G_F^2 E_{\nu}^5 (2\delta_{\nu e})^3}{192\,\pi^3} = 1.3 \times 10^{-14}E_{GeV}^5\delta_{\nu e}^3 \ \ {\rm GeV}
\end{equation}
The mean decay time is then just 1/$\Gamma$. To obtain the numerical value of the mean decay time for VPE, we note that in units where $\hbar = 1$, 1 GeV = $6.58 \times 10^{-25}$ s$^{-1}$.  We adopt the mean fractional energy loss due to a single pair emission of $\sim$ 0.78 from ~\cite{cg11}.

We  assume for  this
calculation  a flat $\Lambda$CDM  universe with  a Hubble  constant of
H$_0$ = 67.8 km s$^{-1}$ Mpc$^{-1}$, taking $\Omega_{\Lambda}$ = 0.7 and
$\Omega_{m}$  = 0.3.  Therefore, the energy  loss owing to redshifting for  a $\Lambda$CDM universe is given by
\begin{equation}
-(\partial \log  E/\partial t)_{redshift} =  H_{0}\sqrt{\Omega_{m}(1+z)^3 +
  \Omega_{\Lambda}}.
\label{redshift}  
\end{equation}

\section{Calculations of Superluminal Neutrino Propagation}

In order to determine the effect of VPE on putative superluminal neutrinos propagating from cosmological distances we explore a simple example using Monte Carlo techniques to take account of energy losses by both VPE and redshifting. We consider a scenario where the neutrino sources have a redshift distribution that follows that of the star formation rate~\cite{be13}, as appears to be roughly the case for active galactic nuclei and $\gamma$-ray bursts. We assume a source spectrum proportional to $E^{-2}$ between 100 TeV and 100 PeV. We generate 50 million events using these two distributions.  Our final results are normalized to an energy flux of $E_{\nu}^2(dN_{\nu}/dE_{\nu}) \simeq 10^{-8} {\rm GeV}{\rm cm}^{-2}{\rm s}^{-1}{\rm sr}^{-1}$, as is consistent with the IceCube data for both the southern and northern hemisphere for energies between 60 TeV and 2 PeV, particularly when atmospheric charm decay neutrinos~\cite{st79,en08} are included in the background subtraction~\cite{wh14}. In our Monte Carlo runs we consider threshold energies between 1 PeV and 40 PeV for the VPE process, corresponding to values of $\delta_{\nu e}$ between $5.2\times10^{-19} \ {\rm and}~ 3.3 \times 10^{-22}$. By propagating our test neutrinos including energy losses from both VPE and redshifting using a Monte Carlo code, we then obtain final neutrino spectra and compare them with the IceCube results.

\begin{figure}[t]
{\includegraphics[scale=0.58]{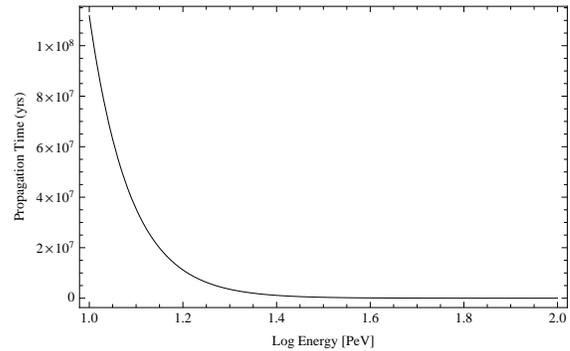}}
\caption{Mean propagation time before decay as a function of neutrino energy for a threshold energy of 10 PeV}
\label{proptime}
\end{figure}

\begin{table}[h]
\centering
\begin{tabular}{|l|l|r|l|l|l|l|}
\hline
Threshold Energy (PeV) & 1 & 2 & 4 & 10 & 20 & 40 \\
\hline
Mean Propagation Time (Gyr) & .011 & .022 & .045 & .11 & .22 & .45 \\
\hline
\end{tabular}
\caption{\label{proptable}Mean propagation time at the threshold energy for the threshold energies considered.}
\end{table}

\section{The IceCube Results}

The IceCube data~\cite{wh14} are plotted in Figure~\ref{spectra}. They are consistent with a spectrum given by $E_{\nu}^2(dN_{\nu}/dE_{\nu}) \simeq \ 10^{-8} \ {\rm GeV}{\rm cm}^{-2}{\rm s}^{-1}{\rm sr}^{-1}$ up to an energy of $\sim$2 PeV, the energy of the so-called "Big Bird" event. No neutrino induced events have been seen above 2 PeV.~\cite{BigBird} 

IceCube has not detected any neutrino induced events from
the Glashow resonance effect. In this effect, electrons in the IceCube volume provide enhanced target cross sections for electron antineutrinos through the $W^-$ resonance channel, $\bar{\nu}_{e} + e^- \rightarrow W^- \rightarrow shower$, at the resonance energy $E_{\bar{\nu}_{e}} = M_W^2/2m_{e} = 6.3$ PeV~\cite{gl60}. This enhancement leads to an increased IceCube effective area for detecting the sum of the ${\nu}_{e}$'s, {\it i.e.}, ${\nu}_{e}$'s plus $\bar{\nu}_{e}$'s by a factor of $\sim 10$~\cite{aa13a}. It is usually expected that 1/3 of the potential 6.3 PeV neutrinos would be ${\nu}_{e}$'s plus $\bar{\nu}_{e}$'s unless new physics is involved~\cite{an13}.
Thus, the enhancement in the overall effective area expected is a factor of $\sim$3. Taking account of the increased effective area between 2 and 6 PeV and a decrease from an assumed neutrino energy spectrum of $E_{\nu}^{-2}$, we would expect about 3 events at the Glashow resonance provided that the number of $\bar{\nu}_{e}$'s is equal to the number of ${\nu}_{e}$'s. Even without considering the Glashow resonance effect, several neutrino events above 2 PeV would be expected if the $E_{\nu}^{-2}$ spectrum extended to higher energies. Thus, the lack of neutrinos above 2 PeV energy and at the 6.3 PeV resonance may be indications of a cutoff in the neutrino spectrum. Hopefully, the acquisition of more data will clarify this point. In the next section we consider the physics implications of both the cutoff and no-cutoff scenarios for the neutrino spectra.

\section{Conclusions and Discussion}

\begin{figure}[t]
{\includegraphics[scale=0.6]{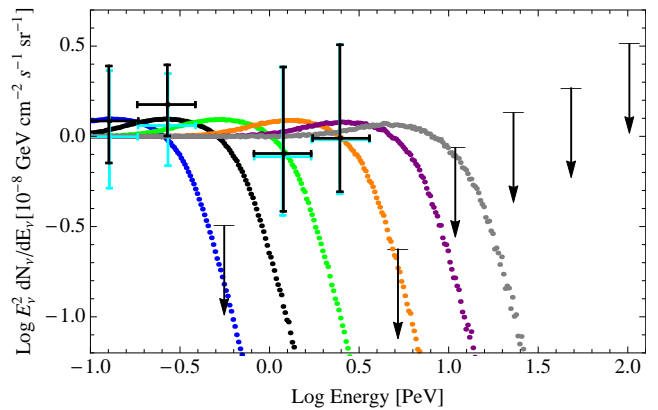}}

\caption{Calculated neutrino spectra with VPE and redshifting compared with the IceCube data both including a subtraction of atmospheric charm $\nu$'s at the 90\% C.L. (cyan) and omitting such a subtraction (black)~\cite{wh14}. Curves from left to right are spectra obtained with rest-frame threshold energies of 1, 2, 4, 10, 20 and 40 PeV. The corresponding values of $\delta_{\nu e}$ are given by equation (\ref{spectra}).}
\label{spectra}
\end{figure}
 
The results of our calculations show that there is a high-energy drop off in the propagated neutrino spectrum resulting from the opening of the VPE channel above threshold. Furthermore, the redshifting effect pushes the cutoff in the energy spectrum below the non-redshifted rest-frame threshold energy. As discussed before, we assume that the neutrino production rate follows the star formation rate in redshift space. This rate peaks at a redshift between 1 and 2. The neutrinos emitted during this past era of enhanced stellar and galactic activity are then redshifted by a factor of 2 to 3. 
The redshifting effect dominates the shape of the resulting spectra regardless of threshold energy.  This is because the mean propagation time is very short compared with the total travel time with the exception of rest-energy thresholds greater than 10 PeV as follows from equations (\ref{threshold}) - (\ref{redshift}) (See table~\ref{proptable}).  Furthermore, the mean propagation time is also short for {\em all} energies greater than the threshold, with the exception of only those very near threshold, as illustrated in Figure \ref{proptime} for a rest-energy threshold of 10 PeV.  In the case of rest-energy thresholds greater than 10 PeV, the particles very near threshold will simply redshift below it without decay.  This has little impact on their final observed energies at $z = 0$.

Our calculated neutrino spectra follow our assumed $E^{-2}$ power-law form below $\sim$0.2 of the the redshifted VPE threshold, have a small pileup effect up to the redshifted threshold energy, and have a sharp high energy cutoff at higher energies, as shown in Figure~\ref{spectra}. The pileup is caused by the propagation of the higher energy neutrinos in energy space down to energies within a factor of $\sim$5 below the threshold. This is indicative the fact that fractional energy loss from the last allowed neutrino decay before the VPE process ceases is 0.78~\cite{cg11}. The pileup effect is similar to that of energy propagation for ultrahigh energy protons near the GZK threshold~\cite{st89}.

Our results yield the best constraints LIV in the neutrino sector to date, {\it viz.}, $\delta_{\nu e} = \delta_{\nu} - \delta_e \le \ 5.2 \times 10^{-21}$. This is because our results for our rest-frame threshold energy cases below 10 PeV as shown in Figure~\ref{spectra} are inconsistent with the IceCube data. Our result for a 10 PeV non-redshifted threshold, corresponding to $\delta_{\nu e} = \ 5.2 \times 10^{-21}$, is just consistent with the IceCube results, giving a cutoff effect above 2 PeV.  We note that the present best upper limit on $\delta_e$ is $5 \times 10^{-21}$~\cite{st14}. Thus for the conservative case of no-LIV effect, {\it e.g.}, if one assumes a cutoff in the intrinsic neutrino spectrum of the sources or one assumes a steeper assumed PeV neutrino spectrum proportional to $E_{\nu}^{-2.3}$~\cite{an13,wh14}, we find the new constraint on superluminal neutrino velocity, $\delta_{\nu} = \delta_{\nu e} + \delta_e  \le \ 1.0 \times 10^{-20}$. However, the steeper spectrum scenario has been placed into question~\cite{ch14}.

Interestingly, for an $E_{\nu}^{-2}$ power-law neutrino spectrum, we find the possibility that
the apparent cutoff in the observed spectrum above $\sim$2 PeV can conceivably be an effect of
Lorentz invariance violation (see Figure~\ref{spectra}). (Another suggestion involving LIV effects of {\it subluminal} neutrinos has recently been discussed~\cite{an14}). A hard $E_{\nu}^{-2}$ spectrum has been proposed to be produced in starburst galaxies~\cite{lo06}. The IceCube flux is below the upper limit of $2 \times 10^{-8} \ {\rm GeV}{\rm cm}^{-2}{\rm s}^{-1}{\rm sr}^{-1}$ obtained by one of us for the neutrino flux from starburst galaxies~\cite{st07}, allowing for this possibility. The power-law source spectrum option opens the possibility that a high energy cutoff in such a hard $E_{\nu}^{-2}$ power-law neutrino spectrum could be caused by a small violation of Lorentz invariance, with  neutrinos being very slightly superluminal, with $\delta_{\nu}$ being $(0.5 \ {\rm to} \ 1.0) \times 10^{-20}$, taking $0 \le \delta_e \le 0.5 \times 10^{-20}$. As has been pointed out previously for ultrahigh energy neutrinos~\cite{go12}, one test for the cutoff scenario would be the non-observation of the  "cosmogenic" neutrinos from photopion production interactions of ultrahigh energy cosmic rays with the cosmic background radiation~\cite{be69}, since all cosmological neutrinos above $\sim$2 PeV would be affected by the VPE process. Such a non-observation would have implications for $\gamma$-ray constraints on ultrahigh energy cosmic ray origin and composition models, perhaps implying the ultrahigh energy cosmic rays are mainly heavy nuclei~\cite{al12}. 

\section*{Acknowledgment} We wish to thank Francis Halzen and Nathan Whitehorn for helpful discussions and information regarding the IceCube results.


\begin{thebibliography}{99}


\bibitem{ad11} T. Adam {\it et al.} (OPERA), arXiv:1109.4897v1.

\bibitem{cg11} A.~G.~Cohen and S.~L.~Glashow, Phys.\ Rev.\ Lett.\ {\bf 107}, 181803 (2011).

\bibitem{ad12} T. Adam {\it et al.} (OPERA), J. High Energy Phys. {\bf 10}, 093 (2012). 

\bibitem{ad13} P. Adamson (MINOS), Nucl. Phys. B (Proc. Suppl.) {\bf 235-236} 296 (2013).

\bibitem{wh14} M. G. Aartsen {\it et al.} (IceCube), arXiv:1405.5303.

\bibitem{aa13a} M. G. Aartsen {\it et al.} (IceCube), Phys. Rev. Lett. {\bf 111}, 021103,
%\bibitem{aa13b} 
M. G. Aartsen {\it et al.} (IceCube), Science {\bf 342}, 1242856 (2013),
%\bibitem{la13} 
R. Laha, J. F. Beacom, B. Dasgupta, S. Horiuchi, and K. Murase, Phys. Rev. D {\bf 88}, 043009.

\bibitem{st14} F. W. Stecker, Astropart. Phys. {\bf 56}, 16 (2014).

\bibitem{bo13} E. Borriello, S. Chakraborty, A. Mirizzi, and P. D. Serpico, Phys. Rev. D {\bf 87} 116009 (2013),
%\bibitem{ma14} 
D. Maz\'{o}n, Phys. Rev. D {\bf 89}, 056012 (2014), J. S. D\'{i}az, V. A. Kosteleck\'{y} and M. Mewes, Phys. Rev. D {\bf 89} 043005 (2014).

%\bibitem{ha13} F. Halzen, {\it Proc. 33rd Intl. Cosmic Ray Conf., Rio de Janeiro}, 2013.

\bibitem{ra13a} S. Razzaque, Phys. Rev. D {\bf 88}, 081302 (2013),
%\bibitem{ah13} 
M. Ahlers and K. Murase, arXiv:1309.4077.

\bibitem{wi13} W. Winter,  Phys. Rev. D {\bf 88}, 083007 (2013),
%\bibitem{ch13} 
I. Cholis and D. Hooper, JCAP {\bf 06} 030 (2013),
%\bibitem{ki13} 
M. D. Kistler, T. Stanev and H. Yuksel, arXiv:1301.1703.

\bibitem{st91} F. W. Stecker, C. Done, M.H. Salamon and P. Sommers,
Phys. Rev. Lett. 66, 2697 (1991), 
J. Alvarez-Mu\~{n}iz and P. M\'{e}sz\'{a}ros, Phys. Rev. D {\bf 70}, 123001 (2004),
F. W. Stecker, Phys. Rev. D {\bf 88}, 047301 (2013).

\bibitem{li13} R.-Y. Liu and X.-Y. Wang, Astrophys. J. {\bf 766}, 73 (2013),
%\bibitem{ra13b} 
Razzaque, Phys. Rev. D {\bf 88}, 103003 (2013), K. Murase and K. Ioka, Phys. Rev. Letters {\bf 111},
121102 (2013).

\bibitem{ka13} O. E. Kalashev, A. Kusenko and W. Essey, Phys. Rev. Letters {\bf 111}, 041103 (2013).

\bibitem{mu13} K. Murase, M. Ahlers and B. C. Lacki, Phys. Rev. D {\bf 88}, 121301 (2013),
%\bibitem{mu14} 
K. Murase, Y. Inoue and C. D. Dermer, arXiv:1403.4089.

\bibitem{st79} F. W. Stecker, Astrophys. J. {\bf 228}, 919 (1979).

\bibitem{ck98} D. Calladay and V. A. Kosteleck\'{y}, Phys. Rev. D {\bf 58}, 116002 (1998).

\bibitem{st92} F. W. Stecker,O. C. de Jager and M. H. Salamon, Astrophys. J. Letters {\bf 390}, L49
(1992).

\bibitem{be13} P. S. Behroozi, R. H. Wechsler and C. Conroy, Astrophys. J. {\bf 770}:57 (2013).

\bibitem{co99} S. R. Coleman and S. L. Glashow, Phys. Rev. D {\bf 59}, 
116008 (1999).

\bibitem{sg01} F. W. Stecker and S. L. Glashow, Astropart. Phys. {\bf 16}, 97 (2001).

\bibitem{en08} R. Enberg, M. H. Reno and I. Sarcevic, Phys. Rev. D {\bf 78}, 043005.

\bibitem{BigBird} We have presented evidence that the cosmic IceCube neutrino events are extragalactic. We note that the 2 PeV event comes from a direction near the galactic plane. If this event is galactic, given that the 1 PeV "Ernie" event is well off the galactic plane and definitely extragalactic, the appropriate curve in Figure 3 would be the 4 PeV threshold curve, corresponding to $\delta_{\nu e} = 3.3 \times 10^{-20}$ and $\delta_{\nu} \le 3.8 \times 10^{-20}$.

\bibitem{gl60} S. L. Glashow, Phys. Rev. {\bf 118}, 316 (1960).

\bibitem{an13} L. A. Anchordoqui, V. Barger, I. Cholis, H. Goldberg, D. Hooper, A. Kusenko, J. G. Learned, D. Marfatia, S. Pakvasa, T. C. Paul and T. J. Weiler, arXiv:1312.6587.

\bibitem{st89} F. W. Stecker, Nature {\bf 342}, 401 (1989).

\bibitem{ch14} C-Y. Chen, P. S. Bhupal Dev and A. Soni, Phys. Rev. D {\bf 89}, 033012 (2014).

\bibitem{an14} L. A. Anchordoqui, V. Barger, H. Goldberg, J.G. Learned, D. Marfatia, S. Pakvasa, T.C. Paul, and T.J. Weiler, arXiv:1404.0622.

\bibitem{lo06} A. Loeb and E. Waxman, JCAP 05 (2006) 003.

\bibitem{st07} F. W. Stecker, Astropart. Phys. {\bf 26}, 398 (2007).

\bibitem{go12} P. W. Gorham, A. Connolly, P. Allison, J. J. Beatty, K. Belov, D. Z. Besson, W. R. Binns, P. Chen, J. M. Clem, S. Hoover, {\it et al.}, Phys. Rev. D {\bf 86}, 103006 (2012).

\bibitem{be69} V. S. Berezinsky and G. T. Zatsepin, Phys. Letters {\bf 28B}, 423 (1969),
%\bibitem{st73} 
F. W. Stecker, Astrophys. and Space sci. {\bf 20}, 47 (1973).

\bibitem{al12} D. Allard, M. Ave, N. Busca, M. A. Malkan, A. V. Olinto, E. Parizot, F. W. Stecker and T. Yamamoto, J. Cosmol. and Astropart. Phys. 09 (2006) 005, G. B. Gelmini, O. Kalashev and D. V. Semikoz, J. Cosmol. and Astropart. Phys. 01 (2012) 044.
\end{thebibliography}
\end{document}